\documentclass{article}

\usepackage{arxiv}

\usepackage[utf8]{inputenc} 
\usepackage[T1]{fontenc}    
\usepackage{url}            
\usepackage{booktabs}       
\usepackage{nicefrac}       
\usepackage{microtype}      
\usepackage{lipsum}		
\usepackage{graphicx}
\usepackage{natbib}
\usepackage{doi}
\usepackage{amsmath,amsfonts,amssymb}
\usepackage{setspace}
\usepackage{tocloft}
\usepackage{float}
\usepackage[table]{xcolor}
\usepackage{macro_ref}
\usepackage{caption}
\usepackage{subcaption}
\usepackage{lineno}
\usepackage{soul}

\title{Asgard/NOTT: L-band nulling interferometry at the VLTI\\II. Warm optical design and injection system}

\author{}

\date{}

\renewcommand{\shorttitle}{Garreau et al.: Asgard/NOTT: L-band nulling interferometry at the VLTI. II.}

\hypersetup{
pdftitle={AsgardNOTT: L-band nulling interferometry at the VLTI. II. Warm optical design and injection system},
pdfsubject={astro-ph.IM},
pdfauthor={Germain Garreau},
pdfkeywords={instrumentation; integrated optics; exoplanets; nulling interferometry; Very Large Telescope Interferometer; Asgard/NOTT},
}

\begin{document}
\maketitle
\vspace{-6em}
\begin{center}
Germain Garreau$^{a}$\footnote{germain.garreau@kuleuven.be},
Azzurra Bigioli$^{a}$,
Romain Laugier$^{a}$,
Gert Raskin$^{a}$,
Johan Morren$^{a}$,
Jean-Philippe Berger$^{b}$,
\newline
Colin Dandumont$^{c}$,
Harry-Dean Kenchington Goldsmith$^{d}$,
Simon Gross$^{e}$,
Michael Ireland$^{d}$,
Lucas Labadie$^{f}$,
\newline
Jérôme Loicq$^{c,g}$,
Stephen Madden$^{d}$,
Guillermo Martin$^{b}$,
Marc-Antoine Martinod$^{a}$,
Alexandra Mazzoli$^{c}$,
\newline
Ahmed Sanny$^{e,f}$,
Hancheng Shao$^{d}$,
Kunlun Yan$^{d}$, and
Denis Defrère$^{a}$
~\newline

$^a$KU Leuven, Institute of Astronomy, Leuven, Belgium\\
$^b$University of Grenoble Alpes/CNRS, IPAG, Grenoble, France\\
$^c$University of Liège, STAR Institute, Liège, Belgium\\
$^d$Australian National University, Research School of Astronomy and Astrophysics, Canberra, Australian Capital Territory, Australia\\
$^e$Macquarie University, School of Mathematical and Physical Sciences, MQ Photonics Research Centre, Macquarie Park, New South Wales, Australia\\
$^f$Universität zu Köln, I. Physikalisches Institut, Köln, Germany\\
$^g$Delft University of Technology, Faculty of Aerospace Engineering, Delft, The Netherlands
\end{center}
~\vspace{-0.5em}

\begin{abstract}
Asgard/NOTT (previously Hi-5) is a European Research Council (ERC)-funded project hosted at KU Leuven and a new visitor instrument for the Very Large Telescope Interferometer (VLTI). Its primary goal is to image the snow line region around young stars using nulling interferometry in the L’-band (3.5 to 4.0)\,µm, where the contrast between exoplanets and their host stars is advantageous. The breakthrough is the use of a photonic beam combiner, which only recently allowed the required theoretical raw contrast of $10^{-3}$ in this spectral range. 
Nulling interferometry observations of exoplanets also require a high degree of balancing between the four pupils of the VLTI in terms of intensity, phase, and polarization. The injection into the beam combiner and the requirements of nulling interferometry are driving the design of the warm optics and the injection system.
The optical design up to the beam combiner is presented. It offers a technical solution to efficiently couple the light from the VLTI into the beam combiner. During the coupling, the objective is to limit throughput losses to 5\,\% of the best expected efficiency for the injection. To achieve this, a list of different loss sources is considered with their respective impact on the injection efficiency. Solutions are also proposed to meet the requirements on beam balancing for intensity, phase, and polarization.
The different properties of the design are listed, including the optics used, their alignment and tolerances, and their impact on the instrumental performances in terms of throughput and null depth. The performance evaluation gives an expected throughput loss $<$6.4\,\% of the best efficiency for the injection and a null depth of $\sim$2.$10^{-3}$, mainly from optical path delay errors outside the scope of this work.
\end{abstract}

\keywords{instrumentation; integrated optics; exoplanets; nulling interferometry; Very Large Telescope Interferometer; Asgard/NOTT}

\newpage

\section{Introduction}\label{sec:intro} 

Imaging planetary systems close to the snow line is an important step in astronomy. This region is known to have a statistically rich exoplanet distribution \citep{fulton_california_2021}. Long baseline interferometry already manages to probe exoplanets using high angular resolution ($\sim$50\,mas) and high contrast ($\sim 10^{-5}$) with the VLTI/GRAVITY instrument \citep{gravity_collaboration_first_2019} at the Very Large Telescope Interferometer (VLTI). 
However, to go below the 50\,mas and image inside the coronagraphic regime \citep{Serabyn2022}, a nulling interferometer becomes advantageous in the mid-infrared range. 
Nulling interferometry was first proposed by Bracewell \citep{bracewell_detecting_1978} and has already been implemented for example with the Palomar Fiber Nuller (PFN) \citep{serabyn_nulling_2019}, the Keck Interferometer Nuller \citep{colavita_keck_2009}, the Nulling and Imaging Camera (NIC) \citep{hinz_nic_2008} at the Large Binocular Telescope (LBT) and the Guided-Light Interferometric Nulling Technology \citep{Norris2020,martinod_scalable_2021} (GLINT) on the Subaru telescope.
Asgard/NOTT (previously Hi-5) is a nulling interferometer \citep{defrere_hi-5_2018,defrere2022} dedicated to high-contrast imaging. It is part of the Asgard instrumental suite \citep{Martinod2023} which is the new visitor instrument of the VLTI.
The instrument would be the first to implement nulling interferometry at the VLTI, aiming to achieve a contrast of $10^{-5}$ after post-processing.

The instrument will use an integrated optics beam combiner to recombine the light from the four Unit Telescopes (UTs) or Auxiliary Telescopes (ATs). Photonic beam combiners have already been successfully implemented at shorter wavelengths in other interferometric instruments like in VLTI/PIONIER \citep{berger_pionier_2010,LeBouquin2011} or Subaru/GLINT \citep{Norris2020,martinod_scalable_2021}. It enables a more stable and more compact design as well as an easier alignment procedure than with free-space optics.
The innovation brought by Asgard/NOTT is the use of such integrated optics in the L'-band (3.5 to 4.0)\,µm \citep{Sanny2022}, a spectral region where guided optics is still emerging. 

The optical design of the instrument includes the cold optics located inside a cryostat and cooled at 90K \citep{Dandumont2022spectro}. They comprise the injection system, the photonic chip, the spectrograph, and the detector. Upstream of the cryostat are the warm optics at ambient temperature. This paper focuses on the warm optics and the injection system up to the waveguides' inputs of the photonic chip.
The coupling between free-space beams with imperfect wavefronts and the guided modes inside a photonic chip requires careful engineering for the optical design. On top of this challenge, nulling interferometry is also particularly demanding when it comes to the balancing of the different beams in terms of intensity, phase, and polarization.

This work presents the warm optical design and the injection system of Asgard/NOTT up to the photonic beam combiner. The proposed design offers a technical solution to maximise the coupling of the light inside the guided modes while enhancing the contrast achievable with nulling interferometry. Section\,\ref{sec:instrumental requirements} first introduces the instrumental requirements which drive the overall optical design. In Sec.\,\ref{sec:design}, a complete description of the optical design is given. Finally, Sec.\,\ref{sec:tolerance analysis} estimates its performances in terms of null depth and coupling efficiency between the light and the guided modes of the beam combiner.

\section{Instrumental Requirements}\label{sec:instrumental requirements}
In this section, we define the objective as well as the requirements for the optical design of the instrument.

\subsection{Requirements on Coupling Efficiency}\label{sec:Description of the photonic chip}

\begin{figure}[H]
    \centering
    \includegraphics[height = 9cm]{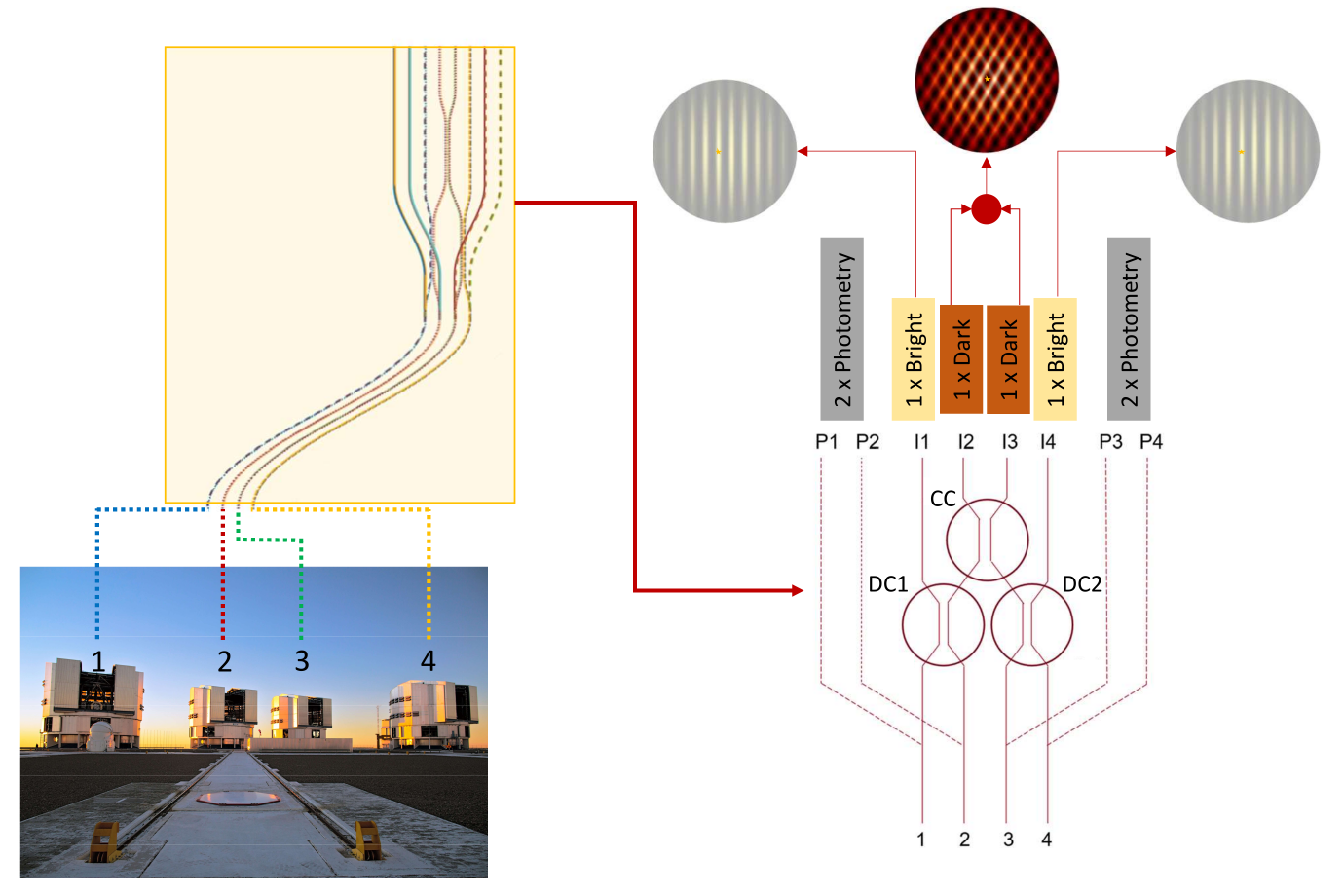}
    \caption{Schematic layout of the photonic chip designed for Asgard/NOTT. On the left, the reproduction of the entire chip with inputs coming from the VLTI telescopes. The S-bend used to reduce stray light after the input is visible. On the right, a closer view of the couplers and the eight outputs. The photometric outputs P1, P2, P3, and P4 emit 20\,\% of the injected light. The bright interferometric outputs I1 and I4 mainly emit the on-axis stellar signal. The dark interferometric outputs I2 and I3 emit the off-axis signal corresponding to possible exoplanets or disks. Source: Ref. \citep{Sanny2022}.}
    \label{fig:chip schema}
\end{figure}

The studied chip beam combiner (or ``chip") is an integrated optic fabricated from a dielectric substrate where an ultrafast laser inscription (ULI) changed the refractive index of the material locally to create optical waveguides \citep{labadie_photonics-based_2020}. The use of couplers inside the chip enables to generate interferences by evanescent coupling between the different waveguides (see Fig.\,\ref{fig:chip schema})\,. The chip planned to be used in the Asgard/NOTT instrument is manufactured at Macquarie University. It is made of gallium lanthanum sulfide (GLS) chalcogenide glass with propagation losses as low as 0.22\,dB/cm at 4\,µm wavelength \citep{gretzinger_towards_2019}. This chip has four inputs corresponding to the four telescopes of the VLTI with the following properties for each input:
\begin{equation}\label{eq:chip input parameters}
     \left\{
\begin{array}{c}
     \text{Numerical Aperture (NA)}=0.1 \\
     \text{Core physical diameter}= 19.4\times19.8\,\text{µm} \\
     \text{$1/e^2$ Mode Field Diameter (MFD)}= 22.3\pm0.1\times22.6\pm0.1\,\text{µm\,@$\lambda$=4\,µm}\\
     \text{Input Separation}= 125\,\text{µm}
\end{array}
\right..
\end{equation}

The performance evaluation in Sec.\,\ref{sec:tolerance analysis} considers a circular MFD of 22.45\,µm. The impact of the MFD error and ellipticity is estimated in Sec.\,\ref{sec:MFD error and ellipticity}. The eight outputs of the chip shown in Fig.\,\ref{fig:chip schema} are spectrally dispersed using grisms with spectral resolutions of 20, 400, or 2000. They are then imaged on a HAWAII 2RG® detector array \citep{Dandumont2022spectro}. Different substrates were available for the chip, for example with lithium niobate \citep{Heidmann2011,Martin2014,Martin2014b} or silicon. These alternatives have benefits: internal fringe locking for lithium niobate or propagation losses $<<$1\,dB/cm for silicon. They may be tested as future prospects using the optical design described in this paper.

The coupling efficiency $\rho$ corresponds to the energy ratio between the incoming electric field and the light that actually couples and propagates into the waveguide. When simplifying the problem by considering a Gaussian approximation for the single-mode waveguide and an Airy pattern from a diffraction-limited regime for the incoming electric field, a simple expression for the coupling efficiency is obtained \citep{ruilier1999} with
\begin{align}\label{eq:coupling efficiency}
    \rho = 2 &\left[\frac{e^{-\beta^2}-e^{-\beta^2\alpha^2}}{\beta\sqrt{1-\alpha^2}}\right]^2, \\
    \label{eq:beta coupling efficiency}
    \text{and }&\beta = \frac{\pi}{2}\frac{\omega_0}{\lambda}\frac{1}{f_\#},
\end{align}
with $\alpha$ the central obstruction, $\omega_0$ the mode radius of the waveguide, $\lambda$ the wavelength of the beam and $f_\#$ the f-number of the injection optic. The resolution of this equation with no obstruction ($\alpha=0$) gives a maximum value of $\rho \simeq 81\,\%$ for $\beta=1.1209$, which corresponds to $f_\#\simeq3.93$ @$\lambda$=4\,µm according to the MFD given in Eq.\,(\ref{eq:chip input parameters}).

One main objective of the design presented in this paper is to maximize the coupling efficiency of each beam into the photonic chip.
This value is very sensitive to optical aberrations, alignment errors, vignetting, phase wavefront perturbations, diffraction, and manufacturing errors of the optics. Some optical aberrations are inherent to the optical design and cannot be minimized or corrected without adding other surfaces to the design. This is the case for instance of coma produced by off-axis parabolas (OAPs). In this paper, the maximum coupling efficiency is defined as the efficiency obtained in a situation where only irreducible aberrations are considered (see Sec.\,\ref{sec:optimal performance}). The requirement is therefore to limit the impact from all other sources at $5\,\%$ of the maximum coupling efficiency, assuming a perfect wavefront at the entrance. The number of 5\,\% is chosen arbitrarily to be less than the losses from the wavefront correction upstream (see Sec.\,\ref{sec:surface errors impact}). Section\,\ref{sec:tolerance analysis} lists the estimated impact from individual sources and the resulting throughput loss that can be expected. The Fresnel losses of the system are also estimated but they are not impacting the coupling efficiency. Therefore, they are not accounted in the 5\,\% criterion.

\subsection{Requirements on the Null Depth}\label{sec:null depth requirements}

\subsubsection{Null depth definition}
The null depth of the instrument corresponds to the ratio between the stellar flux transmitted to the null output (or stellar leakage) and the total stellar flux collected by the telescopes. A lower null depth enables the detection of fainter companions around the host star. To achieve its science goals, Asgard/NOTT targets a raw null depth of $N_{\text{raw}}\leq 10^{-3}$ and a calibrated null depth of $N_{\text{calibrated}}\leq10^{-5}$ over the full L'-band \citep{defrere_hi-5_2018}. These values are very sensitive to imbalances between the different beams regarding intensity, polarization and phase. To achieve $N_{\text{raw}}\leq 10^{-3}$, the sources of errors should each aim for a null depth $\leq 10^{-4}$ as a safe margin.

\subsubsection{Impact of the phase error}\label{sec:Phase requirements}
The relative phase difference between the four beams may have various sources: mainly the Optical Path Difference (OPD) and the longitudinal dispersion from water vapor \citep{absil2006} and CO$_2$.
The OPD represents the difference of optical path $\Delta x$ between two light beams. The associated phase difference $\Delta\phi$ scales with the optical frequency $\nu$ as
\begin{equation}\label{eq:Phase achroma}
    \Delta\phi = 2\pi n \nu \frac{\Delta x}{c} = 2\pi n \frac{\Delta x}{\lambda},
\end{equation}
with c, the speed of light in vacuum and n, the refractive index of the medium of propagation. 
After correction of the static phase difference, the total relative phase fluctuation $\sigma_\phi$ is limiting the null depth $N_\phi$ \citep{serabyn_nulling_2000}

\begin{equation}\label{eq:impact phase on null}
    N_{\phi} > \frac{\sigma_\phi^2}{4}.
\end{equation}

\begin{figure}[H]
    \centering
    \includegraphics[width=\linewidth]{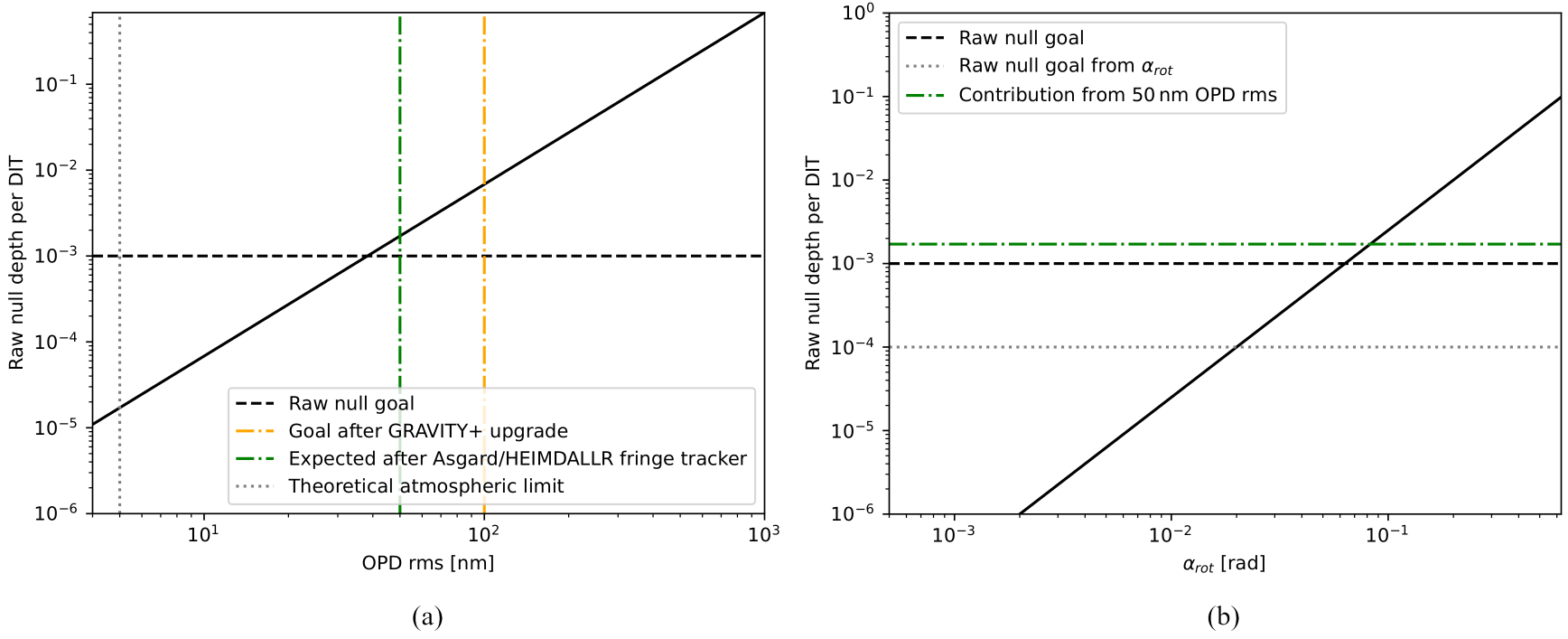}
    \caption{(a) Impact of OPD\,RMS errors on the raw null depth for the UTs. The effects of OPD\,RMS correction from GRAVITY+ and Asgard/HEIMDALLR are given with the two vertical lines. (b) Impact of the polarization angle error on the raw null depth and comparison with the impact of phase error from OPD\,RMS after correction by Asgard/HEIMDALLR.}
    \label{fig:pol&hi5_contrast}
\end{figure}

Current levels of OPD errors with the GRAVITY fringe tracker are estimated at 200\,nm (0.335\,rad) root mean square (RMS) for the UTs \citep{lacour2019}. Ongoing efforts within the GRAVITY+ consortium aim at improving this value down to 100\,nm (0.168\,rad) RMS \citep{Bigioli2022,Courtney-Barrer2022}. The adaptive optics of Asgard/HEIMDALLR \citep{ireland_image-plane_2018} are expected to reduce it down to 50\,nm (0.084\,rad) RMS \citep{Martinache2018}. Using Eqs.\,(\ref{eq:Phase achroma}) and (\ref{eq:impact phase on null}), the raw null depth from OPD fluctuations is computed and presented in Fig.\,\ref{fig:pol&hi5_contrast}.

The static OPD is corrected by the internal delay lines of Asgard/NOTT up to $\sim$12\,mm range.

\subsubsection{Impact of the relative intensity error}\label{sec:impact intensity matching on Null}
The relative intensity error between the inputs $\sigma_I$ is limiting the null depth $N_I$ \citep{Martin2014}
\begin{equation}\label{eq:null depth DeltaI}
    N_I>\frac{\sigma_I^2}{16}.
\end{equation}

The aimed relative intensity error is then $\sigma_I\leq4$\,\% for a null depth of $N_I<10^{-4}$ according to Eq.\,(\ref{eq:null depth DeltaI}). Section\,\ref{sec:tolerance analysis} presents the expected intensity error at the injection from propagation through the optical design only. Strategies to control errors from other sources are discussed in Sec.\,\ref{sec:intensity matching}. Their performance evaluation is part of future work.

\subsubsection{Impact of polarization errors}
Relative polarization errors, both static and dynamic, are hampering the null depth of the instrument. In the optical system, this error may arise from a relative polarization rotation angle difference $\alpha_{rot}$ between the beams.
$\alpha_{rot}$ is limiting the null depth of the instrument $N_P$ with \citep{serabyn_nulling_2000}
\begin{equation}\label{eq:Null depth Pol}
    N_P>\frac{\alpha_{rot}^2}{4}.
\end{equation}

This angle should be kept at $\alpha_{rot}\leq 0.02$\,rad in order to have a null depth $N_P<10^{-4}$ according to Eq.\,(\ref{eq:Null depth Pol}) (see Fig.\,\ref{fig:pol&hi5_contrast}). 

Instrumental birefringence may also impact the null depth limit. When light propagates in a birefringent material, it generates a differential TE/TM OPD. If this OPD is different between two beams, that will limit the null depth according to Eq.\,(\ref{eq:impact phase on null}).

An estimation for these two polarization errors is given is Sec.\,\ref{sec:pol error in design} for the optical design and for the VLTI.
We expect only the instrumental birefringence to have a significant impact on the instrumental null depth. The optical design thus includes a correction setup (see Sec.\,\ref{sec:LDC}) to maintain the OPD from birefringence at $<$10\,µm ($<$0.02\,rad), i.e., $<10^{-4}$ null depth.

\section{Optical Design}\label{sec:design}

In this section, we describe the warm optical design, the injection system, and the motivations for their architecture. The optical designs presented in this section were entirely realized using OpticStudio-Zemax. 

\subsection{Objective}

The light from the telescopes of the VLTI reaches the Visitor Instrument table with diameters of 18 mm and separated by 240\,mm. These beams are first expected to go through the Asgard/HEIMDALLR fringe tracker with adaptive optics mirrors reducing the beam diameter down to 12\,mm. For GRAVITY or PIONIER, the injection inside integrated optics is performed using optical fibers \citep{Lippa2018,LeBouquin2011}. However in the context of Asgard/NOTT, optical fibers working in the L'-band that resist cryo-temperatures and also maintain polarization are not mature enough to be implemented here.

The objective is then to use a free-space optical system to inject these 12\,mm-diameter beams into the four inputs of the chip combiner while respecting the requirements on coupling efficiency and null depth stated in Sec.\,\ref{sec:instrumental requirements}.

\subsection{Step-by-step Description of the Main Optics}

\begin{figure}[H]
    \centering
    \includegraphics[height = 18cm]{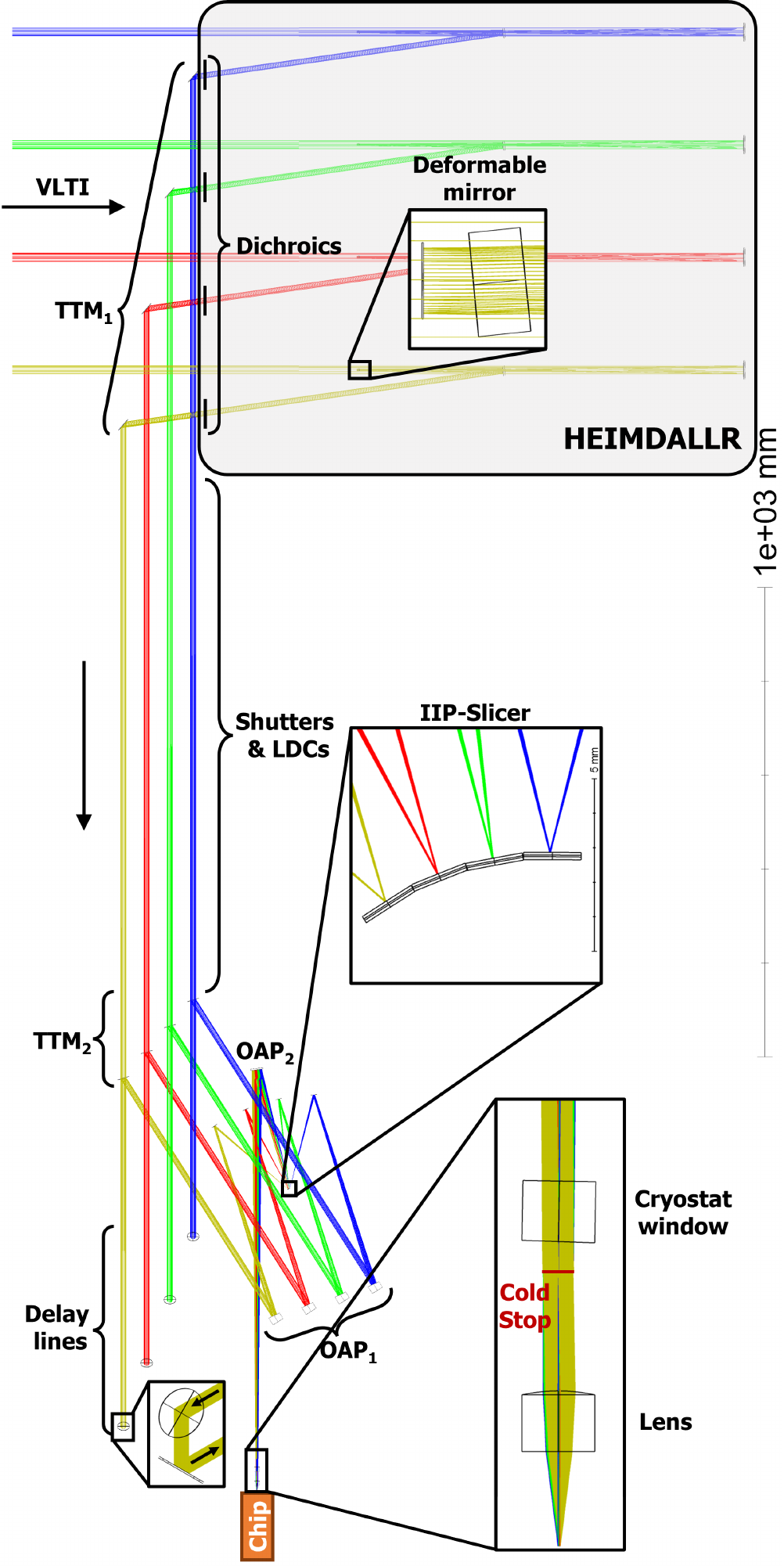}
    \caption{Warm optical design and injection system of the Asgard/NOTT instrument up to the photonic chip. After the dichroics of Asgard/HEIMDALLR, TTMs$_1$ reflect the light throughout the shutters and LDCs. The delay lines then reflect back the beams 40\,mm lower toward the TTMs$_2$. The beams are then focused by the OAPs$_1$ with mirrors reflecting the light on the slicer. The slicer is located in an IIP and recombines the four beams which are then collimated by the OAP$_2$ in the direction of the chip. After traversing the window of the cryostat, the light is injected into the chip using a telecentric lens.}
    \label{fig:injection system}
\end{figure}

Figure\,\ref{fig:injection system} gives a complete view of the optical design.
The four beams from the VLTI are first directed toward the Asgard/HEIMDALLR fringe tracker. Using dichroics optimized to transmit only the L'-band, collimated beams with a diameter of 12\,mm are then reflected along the edge of the optical table. The light beams then encounter shutters, longitudinal dispersion correctors (LDCs) and, the delay lines. These LDCs are presented in Sec.\,\ref{sec:LDC}. The delay lines consist of two plane mirrors with a 45\,$\deg$ angle reflecting the beams in the same direction but 40\,mm lower (going from 125\,mm to 85\,mm height). The two mirrors are moved in parallel to the beams to change their optical path length without affecting the alignment of the system downstream.

Since optical fibers cannot be used to inject into the chip, a free-space optical system is chosen. A multi-lens array like the one used in GLINT \citep{Norris2020} is considered too expensive to be compatible with cryo-temperatures in the L'-band. The solution of a single telecentric lens made of zinc selenide (ZnSe) is then adopted for the injection. 
By implementing a cold stop in the focal plane of the injection lens, the magnification of the image becomes independent from the incidence angle so that mirrors at different angles can be used to recombine the four telescope's pupils.
The use of mirror slices with different angles and located at an intermediate image plane (IIP) enables this recombination; these slices will be assembled in a monolithic optic called ``slicer". OAPs are implemented to focus the incoming beams on the slicer, collimate, and direct them toward the chip and the injection lens. 

Finally, two sets of Tip/Tilt Mirrors (TTMs) are added to the design to ensure the alignment of both the pupils and the images of the beams (see the alignment procedure in Sec.\,\ref{sec:alignment procedure}).

\subsection{Design of the Main Optics}\label{sec: Design of the main optics}

The separation between the slices is chosen to be 1.67\,mm which needs to be demagnified to match the 125\,µm chip's inputs separation. Therefore, the magnification $M$ of the OAP$_2$+lens system needs to be $M=0.075$. At the chip inputs, the beams also need to have $f_{\#\text{, chip}}\simeq3.93$ to match with the result from Eq.\,(\ref{eq:beta coupling efficiency}) to obtain the optimal coupling efficiency with the guided mode of the chip. A pupil diameter D requires an injection lens with a focal length of
\begin{equation}\label{eq:NA to fnumber}
    f_{\text{lens}}=\text{D}\times f_{\#\text{, chip}}.
\end{equation}

The focal length of the OAP$_2$ is then given as 
\begin{equation}\label{eq:focal length M9 from magnification}
    f_{\text{OAP$_2$}} = \frac{f_{\text{lens}}}{M}.
\end{equation}

But the diameter D also constraints $f_{\text{OAP$_2$}}$ from the divergence of the beams at the IIP  $f_{\#\text{, IIP}}$ which gives
\begin{equation}\label{eq:focal length M9 from fnumber}
    f_{\text{OAP$_2$}} = \text{D}\times f_{\#\text{, IIP}}.
\end{equation}

The solution for matching Eqs.\,(\ref{eq:focal length M9 from magnification}) and (\ref{eq:focal length M9 from fnumber}) is then to set
\begin{equation}\label{eq:f-number at slicer}
    f_{\#\text{, IIP}} = \frac{f_{\#\text{, chip}}}{M} .
\end{equation}

$f_{\#\text{, IIP}}$ is defined by the focal length of OAPs$_1$ ($f_{\text{OAP$_1$}}$) and the diameter of the entrance pupil (D$_{\text{en}}$=12\,mm)
\begin{equation}\label{eq:f-number from M5}
    f_{\#\text{, IIP}} = \frac{f_{\text{OAP$_1$}}}{\text{D}_{\text{en}}}.
\end{equation}

From Eq\,(\ref{eq:f-number at slicer}), we can derive the value of $f_\text{OAP$_1$}$
\begin{equation}\label{eq:f-number value}
    f_\text{OAP$_1$} = {\text{D}_{\text{en}}}\times\frac{f_{\#\text{, chip}}}{M} \simeq 629.2\,\text{mm}.
\end{equation}

The only free parameter left for the system is the diameter D of the beam collimated by OAP$_2$. It determines the values for the remaining parameters defined above. D is set to 5\,mm to enable manufacturing and optical alignment easiness, giving the following values:
\begin{equation}\label{eq:optical values}
    D=5\,\text{mm}\Rightarrow
    \left\{
    \begin{array}{c}
        f_{lens}\simeq19.66\,\text{mm} \\
        f_{\text{OAP$_2$}}\simeq262.17\,\text{mm}
    \end{array}
    \right..
\end{equation}



\subsection{Alignment Procedure of the Optical System}\label{sec:alignment procedure}
The alignment of the system is realized by two TTMs for each beam: the first ones (TTMs$_1$) are placed after the dichroics; the seconds (TTMs$_2$) are placed after the delay lines (see Fig.\,\ref{fig:injection system}). These mirrors are used to align the beams both at the pupil position (i.e., at the cold stop) and at the image position (i.e., at the chip inputs). The TTMs$_1$ and TTMs$_2$ are responsible for the alignment at the pupil and image plane respectively. However, since both mirrors are not placed in a conjugate plane, they will have a combined effect in the alignment.

Using a coordinated motion between TTMs$_1$ and TTMs$_2$ (see Sec.\,\ref{sec:TTM angular range}), it is possible to change the alignment of the pupils without impacting the alignment at the chip inputs and vice versa. 
Changing only the pupil alignment with the cold stop may therefore be used to regulate the injected intensity for each beam.
This would enable to actively reduce relative intensity errors mentioned in Sec.\,\ref{sec:impact intensity matching on Null} using the photometric outputs of the chip (see Sec.\,\ref{sec:intensity matching}).


\subsection{Pupil Matching between Asgard/HEIMDALLR and the Cold Stop}\label{sec:pupil matching Plan A}
To maintain a stable injection throughout the cold stop, the alignment of the beams at this position needs to be independent of any tip/tilt generated by Asgard/HEIMDALLR and especially their deformable mirrors (see Fig.\,\ref{fig:injection system}). The solution is to reimage the deformable mirrors at the position of the cold stop. This can be done without altering the light path by imposing convex spherical surfaces to the slicer. With an appropriate radius of curvature, a pupil matching between the cold stop and the deformable mirrors can be achieved, thus providing a stable alignment at the cold stop position.

\subsection{Longitudinal Dispersion Correctors (LDC)}\label{sec:LDC}

\begin{figure}[H]
    \centering
    \includegraphics[height = 8cm]{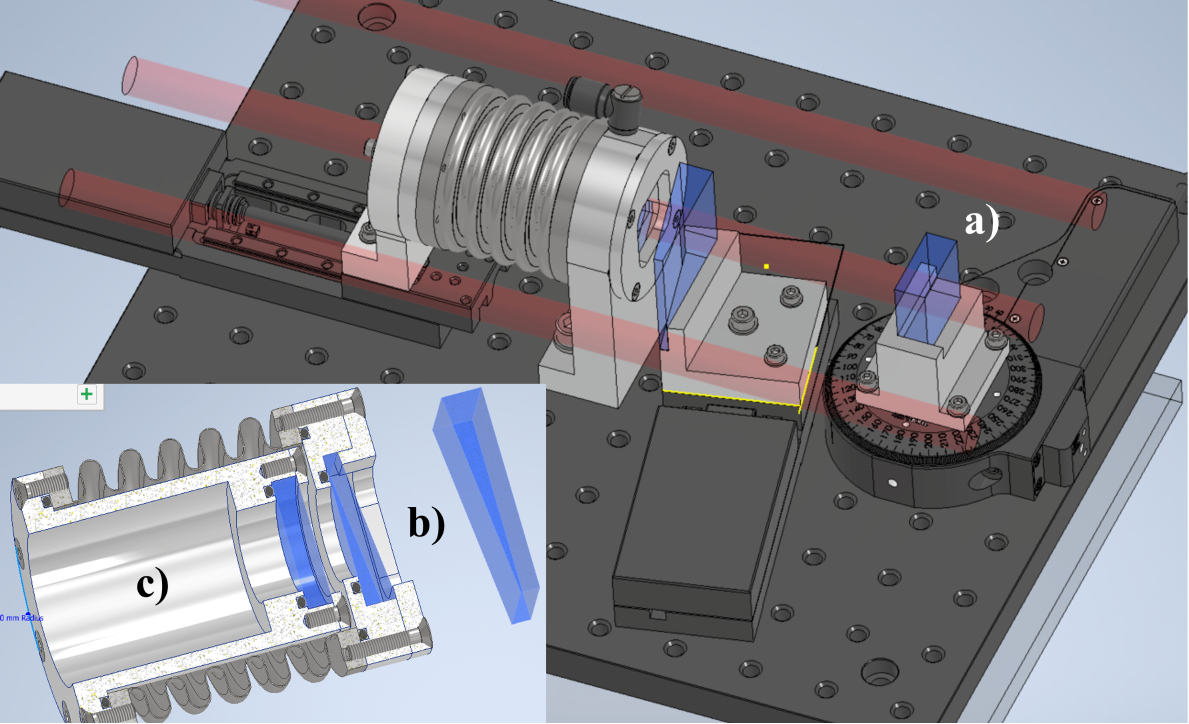}
    \caption{Current opto-mechanical design for the LDC showing a) the birefringence control system, b) the water vapor dispersion control system, 
    and c) the chamber for CO$_2$ dispersion control. The water vapor dispersion control system consists of two LiNbO$_3$ plates: one of them is fixed and used as the window for the chamber, the other plate is on a translation stage. The CO$_2$ chamber is connected to a reservoir to maintain the same pressure of CO$_2$ gas inside while a translation stage moves the left side of the chamber to change its volume.}
    \label{fig:LDC}
\end{figure}

Figure\,\ref{fig:LDC} shows the current opto-mechanical design of the LDCs. The objective of this subsystem is to correct the instrumental birefringence, the water vapor dispersion and the CO$_2$ dispersion altogether for each beam individually.

For the instrumental birefringence, a solution is to use a waveplate and rotate it on a stage until the instrumental birefringence is canceled. The technique was proposed and tested in PIONIER \citep{Lazareff2012}, and achieves a correction down to $<0.1$\,µm OPD precision. However, to achieve a null depth of $10^{-4}$ a correction down to $\sim 10$\,nm is required. A comprehensive study is thus necessary to assess the possibility of reaching such performance.

The correction of water vapor dispersion has already been explored in the case of the Keck Interferometer Nuller \citep{koresko_longitudinal_2003}. A similar design with two wedges of LiNbO$_3$ is planned here: the method is to translate one of the wedges relative to the other to change the light dispersion impacting the transmitted beam. The characteristics of the two wedges can be computed using the performance simulator of Asgard/NOTT: SCIFYsim \citep{Laugier2023}. A wedge angle of 11.5\,deg and a stroke range of 25\,mm are estimated for the current design.

For the CO$_2$ dispersion, its effect on the phase dispersion is observed near and after 4\,µm for example in the VLTI/MATISSE instrument \citep{Lopez2022}. The technical solution proposed in Asgard/NOTT is to use a chamber filled with CO$_2$ at a constant pressure and change its length with a translation stage to control the transmitted beam's dispersion. SCIFYsim enables estimation of the range of the translation stage at 50\,mm.

After correction of the water vapor and CO$_2$ dispersions using the LDCs, the chromatic phase dispersion is estimated to be $<$$0.01$\,rad by SCIFYsim. This corresponds to a null depth of $2.5\times10^{-5}$ according to Eq.\,(\ref{eq:impact phase on null}).

\subsection{Discussion on an Additional IIP for the Optical System}
During the design phase of the instrument, the option of adding a second IIP to the optical system was considered. On paper, this option has one crucial advantage: it would enable the alignment of the pupils at the cold stop and the images independently without combining the movements of two TTMs like in the previously described system.
Indeed, a tip/tilt movement generated at an image plane will change the transversal position of the beam in the pupil plane but not in the image plane. Using the additional IIP as a TTM thus allows an independent alignment and relative intensity error correction at the cold stop. The additional IIP also allows more flexibility on the position of the pupil matching mentioned in Sec.\,\ref{sec:pupil matching Plan A}. Using again convex-shaped spherical surfaces, it is possible to reimage 
the deformable mirrors with the new IIP. At the same time, the slicer can define the entrance pupil position of the cold stop. They can thus be placed at the same position everywhere along the optical axis. If a TTM is then placed at the same location, its movement will only affect the position of the images without disturbing the pupil alignment at the cold stop.

\begin{figure}[H]
    \centering
    \includegraphics[height = 19cm]{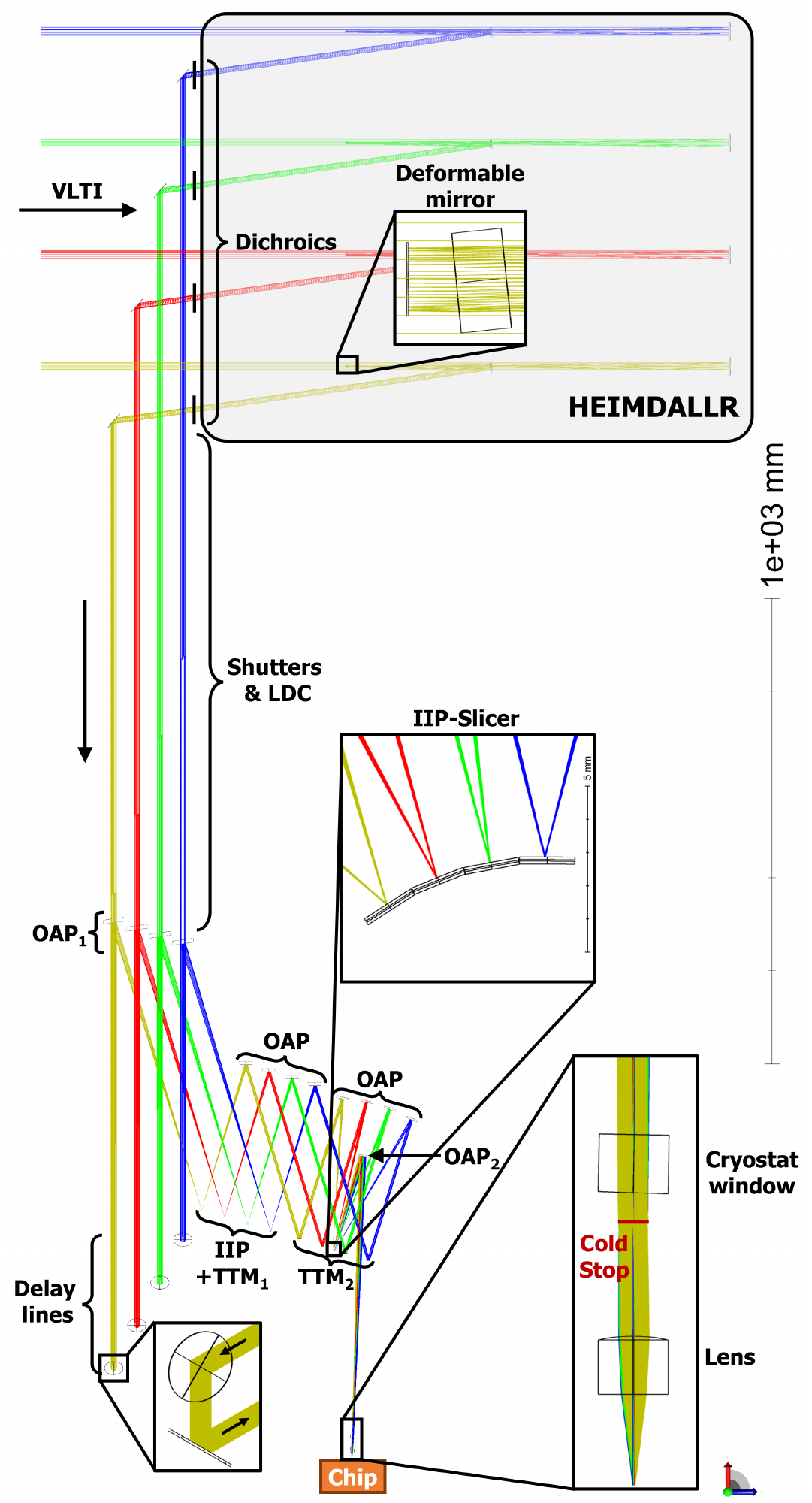}
    \caption{Warm optical design and injection system of the Asgard/NOTT instrument with an additional IIP. After the delay lines, OAPs$_1$ are focusing the beams on TTMs$_1$ located in an IIP. OAPs are then collimating the light toward TTMs$_2$ located at the pupil matching plane. Additional OAPs are focusing the light on a slicer located at an IIP. The slicer is recombining the four beams which are then collimated by the OAP$_2$ in the direction of the chip. After traversing the window of the cryostat, the light is injected into the chip using a telecentric lens.}
    \label{fig:injection system plan B}
\end{figure}

To realize this system, two additional OAPs are needed compared to the optical system in Fig.\,\ref{fig:injection system}. The new IIP is defined at the focus of the OAPs$_1$. The two new OAPs are then used to collimate and refocus the beams on the slicer. A TTM is inserted in between these two OAPs, where pupil matching is done. If the effective focal length of both OAPs is equal, this ensures conservation of the beam divergence. In this case, the optical design in Sec.\,\ref{sec: Design of the main optics} stays valid for the optics downstream of the slicer. A complete view of this system is drawn in Fig.\,\ref{fig:injection system plan B}.

This solution is indeed efficient to perform an independent alignment between the two TTMs. However, its realization implies an increase in cost and complexity. Considering these factors of risk, it has been decided to not pursue this option anymore in the final design.

\section{Performance Evaluation}\label{sec:tolerance analysis}
Using the design presented in Sec.\,\ref{sec:design} and realized in Zemax, this section aims to give a performance evaluation of the system. It is then compared to the Asgard/NOTT requirements listed in Sec.\,\ref{sec:instrumental requirements}.

\subsection{Estimation of the Maximum Coupling Efficiency}\label{sec:optimal performance}

The coupling efficiency is used to evaluate the performance of the injection inside the chip. Using the tools available in Zemax, it is computed for each beam at different wavelengths in the L'-band. When the alignment and the manufacturing of the optics are ideal, the coupling efficiency obtained corresponds to the maximum coupling efficiency of the system as defined in Sec.\,\ref{sec:Description of the photonic chip}. Table\,\ref{tab:optimal Ceff + null cap} shows that a throughput loss of $<$23\,\% (i.e., maximum coupling efficiency of 77\,\%) can be expected during the injection even in an ideal case. This maximum coupling efficiency is different from the one announced by Eq.\,(\ref{eq:coupling efficiency}) ($\sim$81\,\%). The main reason is the use of OAP$_2$ and the injection lens with off-axis beams. This generates aberrations that hamper the coupling efficiency even for perfect alignment and the manufacturing precision.

Table\,\ref{tab:optimal Ceff + null cap} also shows that the injection results in an irreducible relative intensity error even in an ideal case. It translates into a null depth limit from Eq.\,(\ref{eq:null depth DeltaI}) for each wavelength that is well within the requirement from Sec.\,\ref{sec:impact intensity matching on Null}.

\subsection{Tip/tilt Mirrors Evaluation}
\subsubsection{Tip/tilt angular resolution}
The Asgard/NOTT requirement stated in Sec.\,\ref{sec:Description of the photonic chip} mentions that the optical design should provide at least 95\,\% of the maximum coupling efficiencies (see Table\,\ref{tab:optimal Ceff + null cap}). This gives a criterion on the angular resolution needed for the TTMs.
The coupling efficiency computing in Zemax also considers the impact of vignetting upstream from the inputs. The value thus gives a realistic idea of the throughput loss due to misalignments both at the cold stop and chip inputs.

\begin{figure}[H]
    \centering
    \includegraphics[height = 8cm]{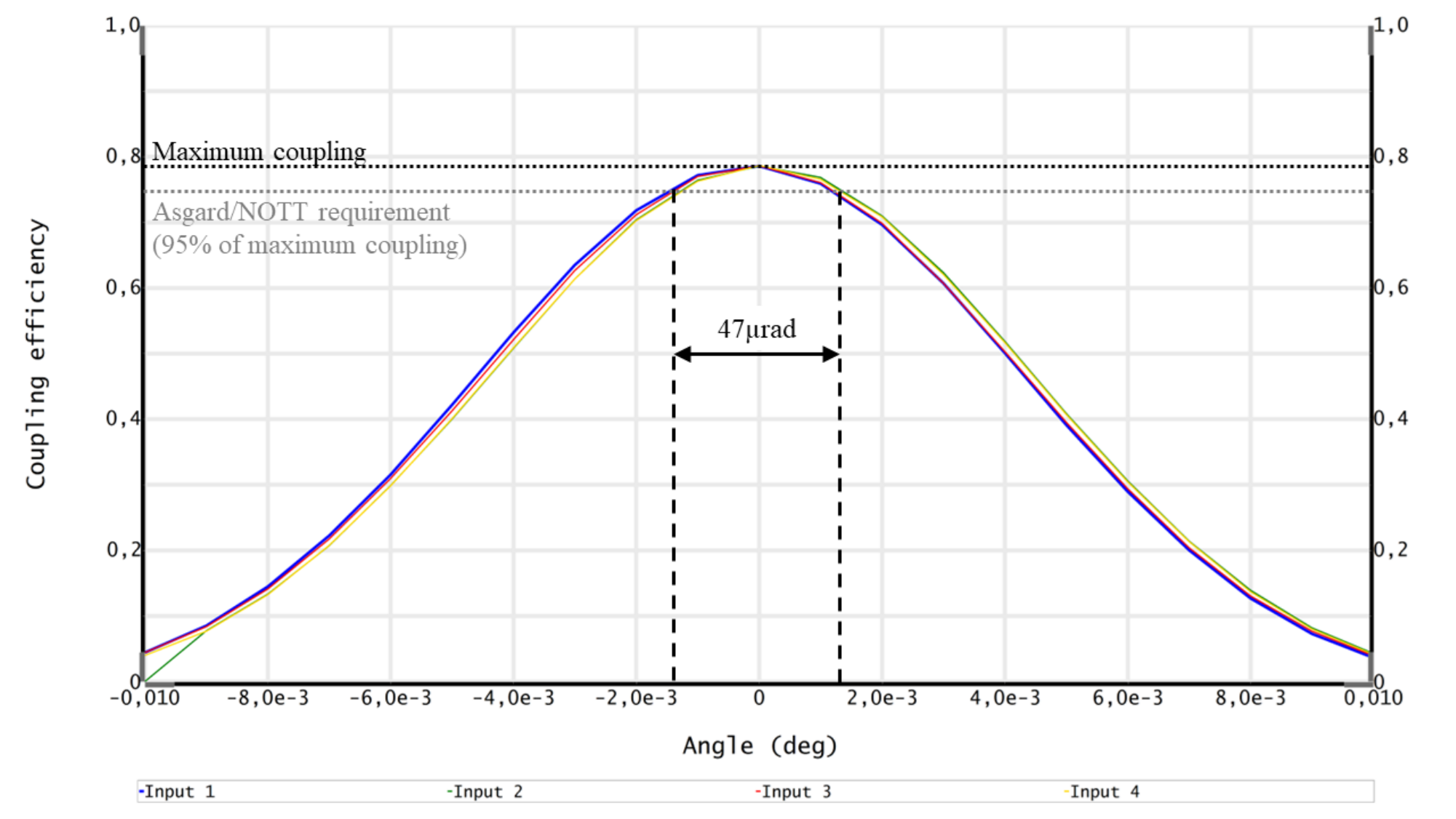}
    \caption{Zemax simulation of the coupling efficiencies in the four chip inputs as a function of the pitch angle from TTMs$_1$. The colors correspond to the beams from Fig.\,\ref{fig:injection system}.}
    \label{fig:TTM1 pitch angle resolution}
\end{figure}

Figure\,\ref{fig:TTM1 pitch angle resolution} shows the sensitivity of coupling efficiency with respect to the pitch angle of TTMs$_1$. Using the 95\,\% criterion defined above, an angle accuracy of 46$\pm$0.4\,µrad is required to stay within this criterion. Assuming a safe margin of 1/10 to find the ideal alignment, an angular resolution of 4\,µrad can be used for this mirror.

Using the same method for the yaw angle and for TTMs$_2$, the results are summarized in Table\,\ref{tab:angular resolution}. Such angular resolution can be reached using piezo optical mounts, for example with Agilis$^{\text{TM}}$ from Newport.

\begin{table}[H]
\caption{Maximum coupling efficiencies computed by Zemax and null depths at 3.5, 3.8 and 4.0\,µm wavelengths. The colors stated for the beams correspond to the colors in Fig.\,\ref{fig:injection system}.}
\begin{center}
\begin{tabular}{|c||c|c|c|}
     \hline
     \cellcolor{black} & \multicolumn{3}{|c|}{\cellcolor{lightgray} Wavelength} \\
     \hline
     \cellcolor{black} Coupling efficiency &\cellcolor{lightgray} 3.5\,µm &\cellcolor{lightgray} 3.8\,µm &\cellcolor{lightgray} 4.0\,µm \\
     \hline \hline
     Input 1 (blue beam) & 76.7\,\% & 78.6\,\% & 77.0\,\% \\
     \hline
     Input 2 (green beam) & 77.0\,\% & 78.6\,\% & 77.1\,\% \\
     \hline
     Input 3 (red beam) & 77.0\,\% & 78.5\,\% & 77.2\,\% \\
     \hline
     Input 4 (yellow beam) & 76.7\,\% & 78.2\,\% & 77.1\,\% \\
     \hline \hline
     \cellcolor{lightgray} Null depth$\times10^{-7}$ & 2.4 & 4.1 & 1.1 \\
     \hline
\end{tabular}
\end{center}
\label{tab:optimal Ceff + null cap}
\end{table}
\begin{table}[H]
    \caption{Simulation results of angular resolution required for the two TTMs to ensure at least 95\,\% of maximum coupling efficiency. A safe margin of about 1/10 is assumed for the final resolution requirement.}
    \begin{center}
    \begin{tabular}{|c||c|c|}
        \hline
         \cellcolor{black} & \cellcolor{lightgray} TTM$_1$ & \cellcolor{lightgray} TTM$_2$ \\
         \hline \hline
         Pitch angle at 95\,\%\,[µrad] & 47$\pm$0.9 & 40$\pm$2.0\\
         \hline
         Yaw angle at 95\,\%\,[µrad] & 46$\pm$1.4 & 38$\pm$1.7\\
         \hline
         \cellcolor{lightgray} Required resolution [µrad] & 4 & 4 \\
         \hline
    \end{tabular}
    \end{center}
    \label{tab:angular resolution}
\end{table}

\subsubsection{Tip/tilt angular range}\label{sec:TTM angular range}
\begin{figure}[H]
    \centering
    \includegraphics[height = 8cm]{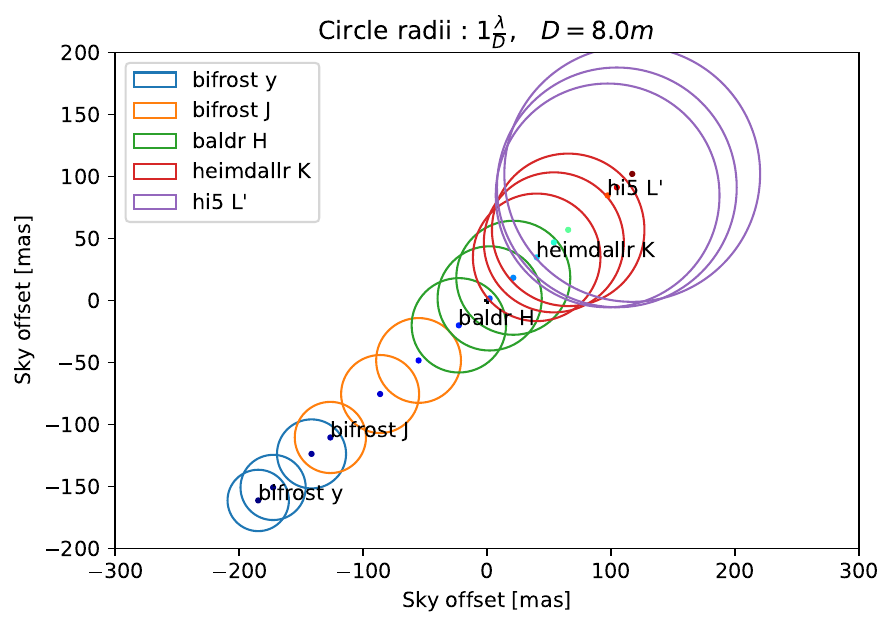}
    \caption{Simulation of the Asgard's field of view projected on-sky with the diffraction pattern's size and position for each instrument (Hi-5 being the older name of NOTT). The simulation corresponds to an observation with the UTs at 30\,deg elevation. The dispersion from Earth's atmosphere shifts the light from the L'-band by an angle of at most 150\,mas compared to the light from the H-band.}
    \label{fig:Asgard FoV}
\end{figure}

The lower limit for the angular range of the TTMs depends on the sky coverage necessary for the instrument. Indeed, the movement from TTMs$_1$ can be used to change the part of the sky injected into the chip inputs, enabling a scan of the field of view. The alignment between the target and the center of the field of view is realized in the H-band by the Asgard/Baldr \citep{Martinod2023} instrument. However, the light dispersion from the atmosphere generates an angular offset between the H-band and the L'-band where Asgard/NOTT operates, as shown in Fig.\,\ref{fig:Asgard FoV} (Laugier et al. in prep.). When observing with the UTs at 30\,deg elevation, this offset is at most 150\,mas projected on-sky. As a safe margin, the system should then be able to have a sky coverage of $\pm$300\,mas around the center of the field of view to ensure a good alignment with the target in the L'-band.

To calculate the resulting physical angle from this offset in the design, the étendue $U$ can be used. $U$ is given by the surface of the pupil $S_{\text{pupil}}$ and the solid angle $\Omega$:
\begin{equation}
    U = S_{\text{pupil}}\times \Omega,
\end{equation}
and remains constant in an optical system \citep{Fischer2008}. The exit pupil of the telescope shares the same étendue as the entrance pupil of the telescope. The physical angle $\theta$ at the exit pupil can thus be calculated from the angle on the sky $\theta_{\text{sky}}$ and the pupil diameters $D_{\text{entrance}}$ and $D_{\text{exit}}$
\begin{align}
    \label{eq:angle at exit pupil for 300mas}
    \theta &= \arcsin{\left(\frac{D_{\text{entrance}}}{D_{\text{exit}}}\times\sin{\left(\theta_{\text{sky}}\right)}\right)}.
\end{align}
The objective is to adapt the system in order to ensure $\theta_{\text{sky}}=\pm$300\,mas in every direction on the sky. For the UTs, $D_{\text{entrance}}=$ 8.2\,m and $D_{\text{exit}}=$ 18\,mm.

\begin{figure}[H]
    \centering
    \includegraphics[height = 4cm]{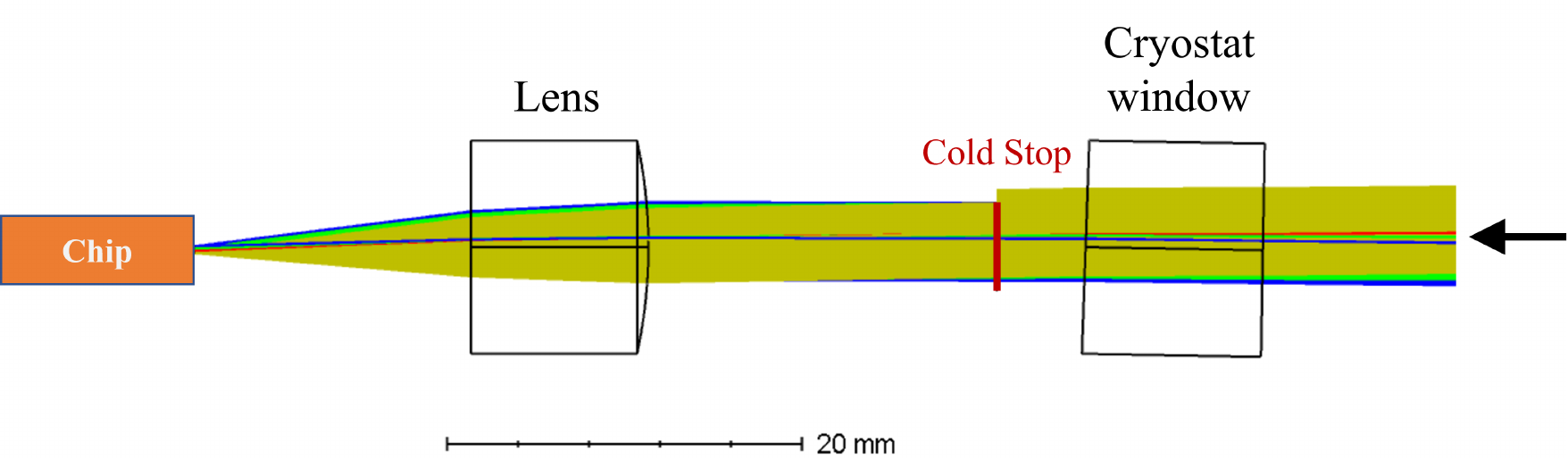}
    \caption{Misalignment at the cold stop when using only the movements of TTMs$_1$ for sky coverage with $\theta_{\text{sky}}=$ 300\,mas. The colors correspond to the beams from Fig.\,\ref{fig:injection system}.}
    \label{fig:cold stop misalignment}
\end{figure}

To compensate for $\theta_{\text{sky}}$, the TTMs$_1$ need to be tilted by an angle $\theta_{\text{TTM$_1$}}=\pm$0.03\,deg. But because TTMs$_1$ are not at the pupil matching position (see Sec.\,\ref{sec:alignment procedure}), this results in a misalignment at the cold stop as shown in Fig.\,\ref{fig:cold stop misalignment}, generating vignetting. The second step is then to use TTMs$_2$ to realign the beams at the cold stop. This can be done with an angle $\theta_{\text{TTM$_2$}}=\pm$0.011\,deg but it also simultaneously moves the spot by $\sim$17\,µm from the chip inputs. The solution is therefore to combine the movements of both TTMs until the four beams are aligned at the chip and the cold stop. Table\,\ref{tab:angular range} summarizes the values for the angles $\theta_{\text{TTM$_1$}}$ and $\theta_{\text{TTM$_2$}}$ for which an optimal alignment is achieved in the case of $\theta_{\text{sky}}=\pm$300\,mas in both directions on the sky. Table\,\ref{tab:angular range} also shows the ratio of vignetted light when using only TTMs$_1$ for the sky coverage. This ratio goes up to $\sim$20\,\% in every direction. The conclusion is that even when only using TTMs$_1$ to search for a target in the field of view, $\sim80\,\%$ of the flux should always be transmitted through the cold stop, enabling the detection of the target.

The angular range required from Table\,\ref{tab:angular range} for both TTMs can be reached using the piezo optical mounts Agilis$^{\text{TM}}$ from Newport for example.

\begin{table}[H]
    \caption{Zemax results of angular range required for the two tip/tilt mirrors. The required range is taken with a safe margin compared to the results of the simulation.}
    \begin{center}
    \begin{tabular}{|c||c|c|c|}
        \hline
         \cellcolor{black} & \cellcolor{lightgray} $\theta_{\text{TTM$_1$}}$\,(µrad) & \cellcolor{lightgray} $\theta_{\text{TTM$_2$}}$\,(µrad) & \cellcolor{lightgray} Vignetting\,(\%) \\
         \hline \hline
         Pitch angle for $\theta_{\text{sky}}$=+300\,mas & -812$\pm$8.7 & -310$\pm$14 & 19$\pm$1.1 \\
         \hline
         Pitch angle for $\theta_{\text{sky}}$=-300\,mas & 812$\pm$8.7 & 314$\pm$17 & 18$\pm$1.0 \\
         \hline
         Yaw angle for $\theta_{\text{sky}}$=+300\,mas & -807$\pm$7.6 & -314 & 18$\pm$0.6 \\
         \hline
         Yaw angle for $\theta_{\text{sky}}$=-300\,mas & 807$\pm$38 & 314$\pm$25 & 19$\pm$0.6 \\
         \hline \hline
         \cellcolor{lightgray} Required angular range [µrad] & $\pm$1000 & $\pm$500 & \cellcolor{black} \\
         \hline
    \end{tabular}
    \end{center}
    \label{tab:angular range}
\end{table}

\subsection{Impact of Fresnel Diffraction}
\subsubsection{Definition}
The Fresnel diffraction model can be used to estimate the encircled energy loss at the pupil plane in the near field. This diffraction may have a significant impact especially when the beam diameter is small ($D=5$\,mm).
The distance along the optical axis z where this approximation is valid is between $z_\text{min}=0.62\sqrt{\frac{(D/2)^3}{\lambda}}$ and $z_\text{max}=\frac{2(D/2)^2}{\lambda}$ \citep{Balanis2005}.

From the Fresnel approximation, the encircled energy $\eta$ is estimated with the relation:
\begin{align}\label{eq:energy from Fresnel}
    \eta(z) &= \frac{\iint_{-D/2}^{+D/2}E(x,y,z)^2dxdy}{\iint_{-\infty}^{+\infty}E(x,y,z)^2dxdy}.\\
    \text{with:\quad} E(x,y,z) &= \frac{e^{ikz}}{i\lambda z}\iint_{-\infty}^{+\infty}E(x',y',0)e^{\frac{ik}{2z}\left[(x-x')^2+(y-y')^2\right]}dx'dy'.
\end{align}
$E(x,y,z)$ represents the electric field at every spatial point and $k$ its wavenumber.

\subsubsection{Results}
\begin{figure}[H]
    \centering
    \includegraphics[height = 6cm]{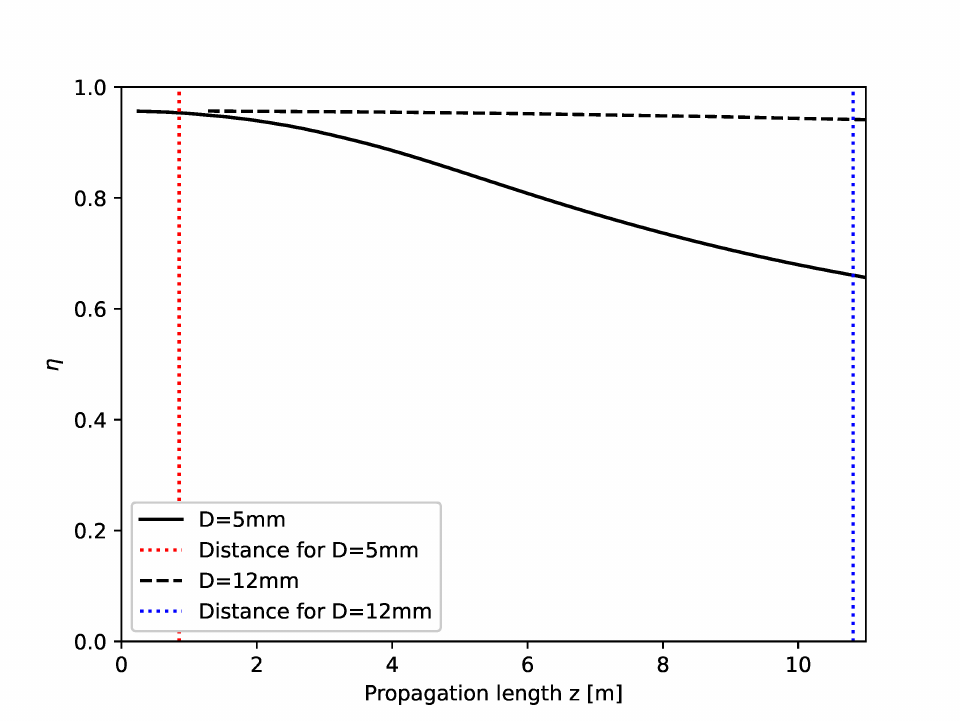}
    \caption{Simulated encircled energy from Fresnel diffraction with respect to propagation length $z$ at $\lambda=4$\,µm. The figure shows the propagation length in the optical design for the beams at 12 and 5\,mm diameter.}
    \label{fig:Fresnel loss}
\end{figure}

To simplify the problem, $\eta$ from Eq.\,(\ref{eq:energy from Fresnel}) is computed in 1D (with $y=0$) at $\lambda=4$\,µm. Figure\,\ref{fig:Fresnel loss} shows the results for the two beam diameters in the system: $D=12$\,mm before OAPs$_1$ and $D=5$\,mm between OAP$_2$ and the injection lens. The results in energy losses for each diameter are presented in Table\,\ref{tab:Fresnel diffraction}. 

\subsection{Impact of Diffraction from the Slicer and the Cold Stop }\label{sec:slicer diffraction analysis}

\begin{figure}[H]
    \centering
    \includegraphics[width = 14cm]{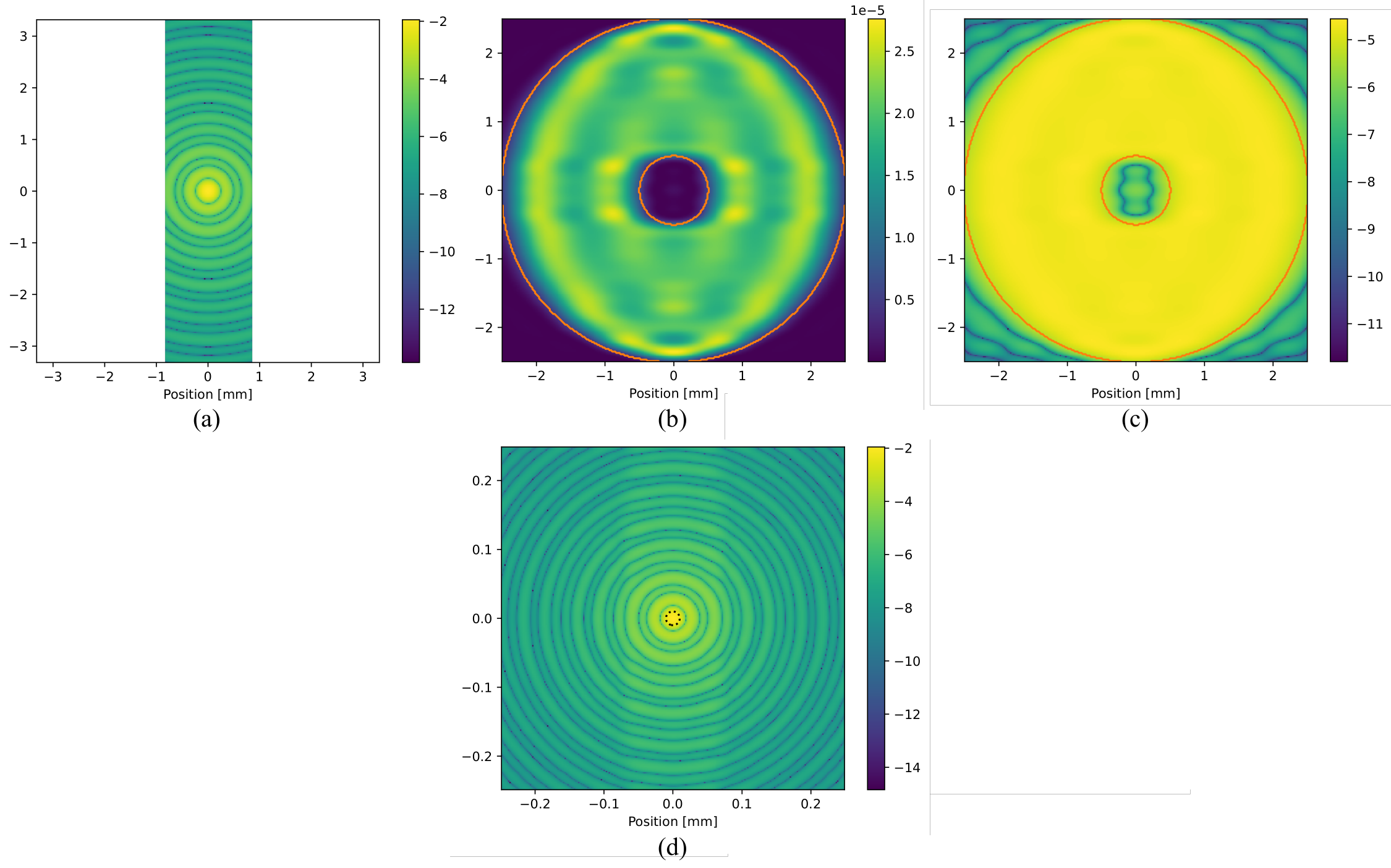}
    \caption{(a) Logarithmic intensity distribution at the slicer for one mirror segment. (b) Linear intensity distribution at the cold stop with diffraction from the slicer segment. The red circles indicate the cold stop diameter and the central obstruction. (c) Logarithmic intensity distribution at the cold stop with diffraction from the slicer segment. The red circles indicate the cold stop diameter and the central obstruction. (d) Logarithmic intensity distribution at the chip's input with diffraction from the slicer segment and the cold stop. The dotted circle indicates the MFD of the chip's waveguide.}
    \label{fig:Diffraction slicer & cold stop}
\end{figure}

The diffraction pattern at the chip's inputs is expected to be altered by the cropping of the intensity distribution at the image plane (from the slicer's segments) and at the pupil plane (from the cold stop). This alteration may impact the maximum coupling efficiency described in Sec.\,\ref{sec:Description of the photonic chip}. Figure\,\ref{fig:Diffraction slicer & cold stop} shows the resulting intensity distribution at the chip's input after the cropping from the slicer's segment and the cropping from the cold stop. Computing the coupling efficiency $\rho$ between the altered diffraction pattern and the MFD gives an estimated loss of 0.08\,\% due to the diffraction from the slicer's segment and the cold stop. This impact is well within our requirement (see Sec.\,\ref{sec:Description of the photonic chip}).

\begin{table}[H]
    \caption{Estimated energy loss $\eta$ from Fresnel diffraction for beam diameters of 12\,mm and 5\,mm at $\lambda=4$\,µm. The propagation length $z$ in both cases is indeed inside the validity range between $z_\text{min}$ and $z_\text{max}$.}
    \begin{center}
    \begin{tabular}{|c||c|c|c|c|}
        \hline
         \cellcolor{black} & \cellcolor{lightgray} $z_\text{min}$\,[m] & \cellcolor{lightgray} $z_\text{max}$\,[m] & \cellcolor{lightgray} $z$\,[m] & \cellcolor{lightgray} $\eta(z)$ \\
         \hline \hline
         D=12\,mm & 0.14 & 18 & 10.8 & 1.5\,\% \\
         \hline
         D=5\,mm & 0.04 & 3.1 & 0.85 & 0.29\,\% \\ \hline\hline
    \end{tabular}
    \end{center}
    \label{tab:Fresnel diffraction}
\end{table}

\subsection{Impact of Mode Field Diameter (MFD) Error and Ellipticity}\label{sec:MFD error and ellipticity}
In this section, the MFD is assumed to be circular with a value of 22.45\,µm. Measurements \citep{gretzinger_towards_2019} indicate an elliptical shape of 22.3x22.6\,µm with an error of $\pm$0.1\,µm. 
The MFD ellipcticity and error may have an impact on the maximum coupling efficiency that can be achieved. This impact is estimated by considering two extreme cases: a circular MFD of 22.2\,µm and 22.7\,µm respectively. In an ideal alignment situation, the worst case scenario is a MFD of 22.2\,µm with a loss of 0.5\,\% on the maximum coupling efficiency at 3.5\,µm wavelength. The impact of waveguide ellipticity is thus assumed to be $<$0.5\,\%, which is well within our requirement (see Sec.\,\ref{sec:Description of the photonic chip}).

\subsection{Fabrication Errors Evaluation}
\subsubsection{Definitions}
We identify two types of fabrication errors that affect the quality of optics: the figure errors and the surface errors. The first one represents the deviation of the full-figure form of the surface from the ideal one. This induces a deviation from the theoretical radius of curvature and conic. The second one represents the local deviations of the surface compared to the full-figure form inducing local wavefront errors.

\subsubsection{Impact of figure errors}
The figure errors on the radius of curvature of a surface are estimated to be less critical as they do not generate aberrations. The change of focal length from these errors can be measured by the manufacturer and later compensated in the optical design. However, this error needs to be constrained as it also changes the value of $NA$ and therefore can impact the performance of the injection into the chip. The tolerances of the curvature for the aspherical optics of the system are summarized in Table\,\ref{tab:figure errors R}. The four OAPs$_1$ are assumed to share the same radius of curvature if they are manufactured from the same parabola.

\subsubsection{Impact of surface errors}\label{sec:surface errors impact}
Surface errors generate wavefront errors which result in aberrations when focusing the beams at the chip inputs. These aberrations are hampering the expected coupling efficiencies. This energy loss from wavefront errors can be approximated by the Strehl ratio.

For small aberrations (normalized wavefront error $\omega\leq1/10$\,RMS), this Strehl ratio S can be estimated using Maréchal's approximation
\begin{equation}
    S\sim e^{-(2\pi\omega)^2},
\end{equation}
with the normalized wavefront error RMS $\omega$ corresponding to
\begin{equation}
    \omega = \frac{\lambda_{\text{ref}}}{N}.\frac{1}{\lambda_0},
\end{equation}
with $\lambda_{\text{ref}}$ the reference wavelength (usually $\lambda_{\text{ref}}=633$\,nm), N the surface quality index and $\lambda_0$ the central wavelength ($\lambda_0=3.75$\,µm in the L'-band).

For a multiple optics combination, the total Strehl ratio of a number of optics $N_{\text{optics}}$ is
\begin{align}
    S&\sim e^{-(2\pi\omega_{t})^2},\label{eq:Strehl ratio} \\
    \text{with:}\quad\omega_{t} &= \sqrt{\sum_{i}^{N_{\text{optics}}}\omega_i^2}.
\end{align}

\begin{table}[H]
    \caption{Simulation results of radius of curvature tolerances for the curved optics to ensure at least 95\,\% of maximum coupling efficiency across the L'-band. A safe margin of 1/10 is assumed for the final tolerances.}
    \begin{center}
    \begin{tabular}{|c||c|c|c|}
        \hline
         \cellcolor{black} & \cellcolor{lightgray} OAPs$_1$ & \cellcolor{lightgray} OAP$_2$ & \cellcolor{lightgray} Lens \\
         \hline \hline
         Tolerances at 95\,\%\,[mm] & $\pm$42 & $\pm$17 & $\pm$0.06 \\
         \hline
         Tolerances at 95\,\%\,[\%] & $\pm$3.4 & $\pm$3.3 & $\pm$0.2 \\
         \hline \hline
         \cellcolor{lightgray} Final tolerances\,[\%] & $\pm$0.3 & $\pm$0.3 & $\pm$0.02 \\
         \hline
    \end{tabular}
    \end{center}
    \label{tab:figure errors R}
\end{table}

For reflective optics, the surface errors $\delta$ and wavefront errors $\omega$ are related by a factor 2
\begin{equation}
    \omega = 2\times \delta.
\end{equation}

For refractive optics, the surface errors $\delta$ and wavefront errors $\omega$ are related by a factor depending on the glass index of refraction n
\begin{equation}
    \omega = (n-1)\times\delta.
\end{equation}

In some cases, the surface errors are given by manufacturers in peak-to-valley (P-V). 
The RMS equivalents are typically approximated to a range going from $\frac{1}{5}$ to $\frac{1}{3}$ of the P-V values \citep{Fischer2008}.

\subsubsection{Estimation of the maximum surface errors}
The objective is to find the maximum surface errors for each optic that reach the objective of a Strehl ratio $S_\text{NOTT}>95$\,\% while minimizing the manufacturing challenges.
In Fig.\,\ref{fig:injection system}, the wavefront errors from optics prior to the dichroics are corrected by the Zernike Wavefront Sensor in Asgard/Baldr using the deformable mirrors of Asgard/HEIMDALLR. Only the optics after and including the dichroics are considered to estimate $S_\text{NOTT}$. This comprises the OAPs$_1$, OAP$_2$ and slicer (three reflective surfaces), the cryostat window and injection lens (four refractive surfaces), the dichroics (two refractive surfaces) and the flat mirrors (five reflective surfaces). This estimation does not include the wavefront errors generated by the LDCs.

\begin{figure}[H]
    \centering
    \includegraphics[height = 8cm]{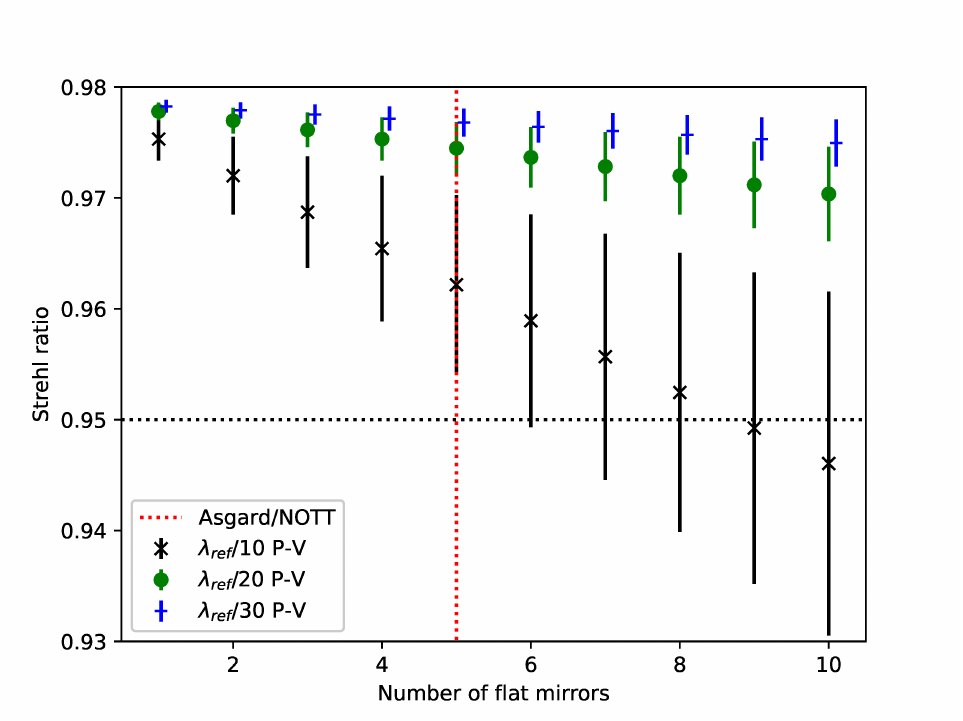}
    \caption{Strehl ratio estimation at the chip input as a function of the number of flat surfaces and their surface quality. The error bars come from the uncertainty in the translation from P-V to RMS values. The estimation is made at 3.75\,µm wavelength with $\lambda_\text{ref}=633nm$.}
    \label{fig:Strehl ratio}
\end{figure}

The OAPs$_1$, OAP$_2$ and slicer are custom optics designed with high surface quality $\left(\frac{\lambda_{\text{ref}}}{40}\,\text{RMS}\right)$. The injection lens is designed with an expected surface quality of $\frac{\lambda_{\text{ref}}}{20}$\,RMS from ZnSe ($n=2.4178$ at $\lambda=3.75$\,µm). The window of the cryostat is chosen with a transmitted wavefront error of $\frac{\lambda_\text{ref}}{10}$\,P-V. Finally the dichroics are made of CaF$_2$ ($n=1.4120$ at $\lambda=3.75$\,µm) with an expected surface quality of $\frac{\lambda_{\text{ref}}}{10}$\,RMS.
These hypotheses allow to study whether the use of commercially available off-the-shelf optics for the flat mirrors is possible.
Figure\,\ref{fig:Strehl ratio} shows that a Strehl ratio $S_\text{NOTT}=97.5\,\%$ is expected with a surface quality of $\frac{\lambda_{\text{ref}}}{20}$ P-V for the flat mirrors. This surface quality is indeed obtainable, e.g., from Edmund Optics. 

The wavefront correction done by Asgard/Baldr is expected to reach a Strehl ratio $S_\text{Baldr}$ of $>$40\.\% and $>$50\,\% for the UTs and the ATs, respectively, in the J-band \citep{Martinod2023}. This corresponds to a Strehl ratio of $>$90.3\,\% and $>$92.6\,\%, respectively, in the L'-band (or a throughput loss of $<$9.7\,\% and $<$7.4\,\%, respectively, during the injection in the chip). Using Eq.\,(\ref{eq:Strehl ratio}), the Strehl ratio expected at the chip's inputs $S$ corresponds to
\begin{equation}
    S=S_\text{Baldr}\times S_\text{NOTT},
\end{equation}
where $S$ is thus estimated to be $>$88.0\,\% and $>$90.3\,\% for the UTs and the ATs respectively in the L'-band.

\subsection{Impact of Polarization Errors}\label{sec:pol error in design}
From the optical design simulated on Zemax, the different polarization errors can be estimated with their impact on the null depth (see Sec.\,\ref{sec:impact intensity matching on Null}). This estimation is done considering three linear polarization states at the beginning of the design: $|V\rangle$, $|H\rangle$, and $|D\rangle$ for ``vertical", ``horizontal", and ``diagonal" polarization states respectively. At the inputs of the chip, the final polarization angle is calculated for the $|V\rangle$ and $|H\rangle$ cases and the difference between the four beams gives a null depth limit using Eq.\,(\ref{eq:Null depth Pol}). In the $|D\rangle$ case, the phase delay between $|V\rangle$ and $|H\rangle$ is also calculated at the inputs of the chip and for each beam. This phase delay is due to the birefringence of the optical system and also contributes to limiting the null depth.
Table\,\ref{tab:pol angle difference} summarizes the results for the three cases. As first expected, birefringence represents the main null depth limitation from polarization errors ($<3.7\times10^{-3}$), which confirms the need for a birefringence correction (see Sec.\,\ref{sec:LDC}). 

For the polarization errors induced by the VLTI optics, a recent study estimated that the differential birefringence between the UTs generates a phase error of $<$$0.0175$\,rad at $2.2$\,µm wavelength \citep{Widmann2023}. This corresponds to an OPD error of $<$$6$\,nm. In the L'-band, the phase error is $<$$0.01$\,rad. This limits the null depth to $2.6\times10^{-5}$, which is within our requirements. For the ATs, larger differential birefringence is expected and can be corrected given the model of its Jones matrix. Other polarization errors (degree of polarization, polarization angle, etc.) may arise from the VLTI optics but a more extensive study is needed to estimate them.

\subsection{Final Throughput Loss and Null Depth Budget}\label{sec:summary on Ceff losses}

This section summarizes the impact of the different sources impairing the optical throughput (Table\,\ref{tab:summary throughput losses}) and the null depth from the optical system (Table\,\ref{tab:summary Null losses}). Table\,\ref{tab:summary Null losses} highlights that OPD errors and birefringence are expected to be the main limitations for the null depth of the instrument. After correction by the LDCs, the null depth from birefringence is expected to be $\leq 10^{-4}$, which leaves the OPD errors as the main limitation (assuming that the null depth is not limited by the photonic chip). This confirms that reducing OPD errors at the VLTI is the best way to improve high-contrast imaging with nulling interferometry.

The impact of alignment and figure errors is calculated with Monte Carlo method on Zemax. Using the values from Table\,\ref{tab:angular resolution} and \ref{tab:figure errors R}, less than 1\,\% throughput loss is expected across the bandwidth and for the four beams.

The throughput loss during the injection is estimated at $<$6.4\,\% of the best coupling efficiency. While it is above the objective, it remains less than the impact from wavefront errors. The optical design is thus assumed to have an optimized injection efficiency. 

An estimation of the Fresnel losses gives a throughput loss of 40\,\%. It represents the losses due to reflections and transmissions from the Asgard/NOTT optics (including the LDCs). The global throughput loss of the instrument is then estimated at $<$60\,\%.

\begin{table}[H]
    \caption{Simulated polarization errors after propagation through the optical system. $|V\rangle$, $|H\rangle$, and $|D\rangle$ represent the ``vertical", ``horizontal", and ``diagonal" linear polarization states at the beginning of the system. For the $|V\rangle$ and $|H\rangle$ cases, the simulation gives the polarization angle difference at the chip's inputs. For the $|D\rangle$ case, the simulation gives the phase delay between $|V\rangle$ and $|H\rangle$ at the chip's inputs. The colors stated for the beams correspond to the color in Fig.\,\ref{fig:injection system}.}
    \begin{center}
    \begin{tabular}{|c||c|c||c|}
         \hline
         \cellcolor{black} & \multicolumn{2}{|c||}{\cellcolor{lightgray} Angle at the chip [mrad]} & \cellcolor{lightgray} Phase delay at the chip [mrad] \\
         \hline
         \cellcolor{lightgray} Starting polarization state  & \cellcolor{lightgray} $|V\rangle$ & \cellcolor{lightgray} $|H\rangle$ & \cellcolor{lightgray} $|D\rangle$ \\
         \hline \hline
         Input 1 (blue beam) & $4.775$ & $4.611$ & $<87.3$ \\
         \hline
         Input 2 (green beam) & $4.761$ & $4.564$ & $<122$ \\
         \hline
         Input 3 (red beam) & $4.753$ & $4.559$ & $<122$ \\
         \hline
         Input 4 (yellow beam) & $4.784$ & $4.625$ & $<87.3$ \\
         \hline \hline
         \cellcolor{lightgray} Null depth & $2.5\times 10^{-10}$ & $10^{-9}$ & $<3.7\times 10^{-3}$ \\
         \hline
    \end{tabular}
    \end{center}
    \label{tab:pol angle difference}
\end{table}

\begin{table}[H]
    \caption{List of phenomena impairing the optical throughput. A first summary is given considering only the phenomena impacting the injection losses that we minimize (see Sec\,\ref{sec:Description of the photonic chip}). A second one including all the combined losses. The Fresnel losses correspond to the reflectivity and the transmittance of the Asgard/NOTT optics (including the LDCs).}
    \begin{center}
    \begin{tabular}{|c||c|}
        \hline
         \cellcolor{black} & \cellcolor{lightgray} Throughput loss [\%] \\
         \hline \hline
         Fresnel diffraction & $1.8$ \\ \hline
         Slicer \& Cold stop diffractions & $0.8$\\ \hline
         MFD ellipticity & $<0.5$\\ \hline
         Alignment \& figure errors & $<1$ \\ \hline
         Surface errors & $2.5$ \\ \hline\hline
         \cellcolor{lightgray} \textbf{Total 1} & \cellcolor{lightgray} $<6.4$ \\
         \hline\hline
         Maximum coupling efficiency & $<23$ \\ \hline
         Wavefront errors (after Asgard/Baldr correction) & $<9.7$\,[UTs] and $<7.4$\,[ATs] \\ \hline
         Fresnel losses & $40$ \\\hline\hline
         \cellcolor{lightgray} \textbf{Total 2} & \cellcolor{lightgray} $<60$ \\ \hline
    \end{tabular}
    \end{center}
    \label{tab:summary throughput losses}
\end{table}

\begin{table}[H]
    \caption{List of phenomena impairing the null depth performance. The birefringence value before correction includes the errors from the UTs and from Asgard/NOTT (the errors from the UTs being negligible). The values corresponding to the intensity error and the birefringence after correction are goals and their estimation is part of future work. The total is calculated assuming the goals for intensity errors and birefringence are met.}
    \begin{center}
    \begin{tabular}{|c||c|}
        \hline
         \cellcolor{black} & \cellcolor{lightgray} Null depth \\
         \hline \hline
         OPD error (after Asgard/HEIMDALLR) & $2\times10^{-3}$ \\ \hline
         Intensity error (objective) & $\leq 10^{-4}$ \\ \hline
         Longitudinal dispersion from H$_2$O \& CO$_2$ & $2.5\times10^{-5}$ \\ \hline
         Polarization angle difference & $10^{-9}$ \\ \hline
         Birefringence before [after] LDCs correction & $<3.7\times 10^{-3}$ [$10^{-4}$]
         \\\hline\hline
         \cellcolor{lightgray} \textbf{Total} & \cellcolor{lightgray} $2.2\times10^{-3}$ \\ \hline
    \end{tabular}
    \end{center}
    \label{tab:summary Null losses}
\end{table}

\section{Discussion}
\subsection{Beam Swapping Capacity}

Because Asgard/NOTT uses a two-stage nulling recombination scheme, its sensitivity map depends on the order in which the telescopes are recombined \citep{Laugier2023}. Therefore, the optical design must enable the swap of the light beams two-by-two before the couplers inside the chip. This swap would allow to ensure more flexibility to detect an exoplanet. 

Several internal solutions have been considered, including beam swapping with different photonic chips, with the slicer, or with mirror permutation, but none of these sufficiently fit the instrumental constraints. An alternative solution is to perform beam swapping using the M16 mirrors of the VLTI and moving the VLTI delay lines accordingly to correct for the OPD. This solution would make Asgard/NOTT simpler and easier to implement at Paranal but also less automatic.

\subsection{Intensity Matching}\label{sec:intensity matching}
Section\,\ref{sec:impact intensity matching on Null} presents the requirement in terms of relative intensity error $\sigma_I$ between the different injected beams. Table\,\ref{tab:optimal Ceff + null cap} shows that the expected contribution from the optical design is well within the null depth requirement of $10^{-4}$. However, other sources of intensity mismatch may contribute, like the difference of throughput between the telescopes of the VLTI. Measurements done in April 2020 show a difference of throughput between the four ATs which gives a relative intensity error of almost 20\,\% in the K-band, i.e., a null depth of $2.5\times 10^{-3}$ (X. Haubois, personal communication, December 14, 2023). This example motivates the need for a strategy to maintain $\sigma_I<4\,\%$.

The intensity mismatch can be directly measured using the photometric outputs of the chip (see Fig.\,\ref{fig:chip schema}).To actively reduce the mismatch, several strategies are considered for the instrument:
\begin{itemize}
    \item Using motorized iris diaphragms, the pupil diameter of each beam can be actively adjusted to minimize $\sigma_I$.
    \item Using the TTMs, misalignments between the pupil planes and the cold stop can generate vignetting to minimize $\sigma_I$ (see Sec.\,\ref{sec:alignment procedure}).
\end{itemize}

However both of these solutions cannot provide a chromatic correction of the intensity injected into the chip. In order to enable a wavelength-dependent control of the intensity as well as the phase, an adaptive nuller \citep{Peters2010} is considered. The system would be a future improvement for the instrument installed in between OAP$_2$ and the cryostat.

\section{Conclusion}
The finalized version of the Asgard/NOTT warm optical design and injection system is presented in this paper. This work gives a complete description of the optical layout as well as explanations for the choices made during the design process. 
The performance section lists different phenomena that can hamper the throughput of the instrument and/or the null depth. The impact for each of these is estimated and an overall null depth limit of $2.2\times 10^{-3}$ can be given for the instrument with a throughput loss of $<$$60\,\%$. The null depth limit is expected to be dominated by the OPD error from the atmosphere and the vibrations at the VLTI, even after correction by the adaptive optics of Asgard/HEIMDALLR. The loss of throughput is dominated by the Fresnel losses and wavefront errors after the Asgard/Baldr correction. For the injection into the chip, the throughput loss is estimated to be $<$6.4\,\% of the maximum coupling efficiency. Even if the estimation is above the requirement of 5\,\% stated in Sec.\,\ref{sec:Description of the photonic chip}, it remains below the loss from wavefront errors. The optical system and its tolerancing are thus assumed to be adapted for efficiently injecting the VLTI beams into the beam combiner of the instrument.

Some aspects of the design are still risky and subject to change in the future like the LDCs for which a prototype demonstration is required. Combining the two TTMs also needs a comprehensive study to allow for alignment and possible intensity matching. These will be implemented in the test bench of the Asgard/NOTT instrument located at KU Leuven.

\subsection*{Acknowledgments} 
SCIFY has received funding from the European Research Council (ERC); Award no. CoG –
866070 under the European Union’s Horizon 2020 research and innovation program. M-A.M. has
received funding from the European Union’s Horizon 2020 Research and Innovation Program;
Award No. 101004719. Acknowledgment to J.B. Le Bouquin (IPAG/OSU) for discussions about
the injection alignment.

\bibliographystyle{unsrtnat}
\bibliography{references}  






\end{document}